\documentclass[prl,aps,10pt,showpacs,twocolumn,unsortedaddress]{revtex4-1}
\usepackage{graphicx}
\usepackage{amsfonts}
\usepackage{amsmath}
\usepackage{amssymb}
\usepackage{bm}

\usepackage[usenames,dvipsnames]{xcolor}
\newcommand{\be}{\begin{equation}}
\newcommand{\ee}{\end{equation}}

\newcommand{\trm}[1]{\textrm{#1}}
\newcommand{\Ecr}{E_{\trm{cr}}}
\newcommand{\figref}[1]{Fig. \ref{#1}}
\newcommand{\eqnref}[1]{Eq. (\ref{#1})}

\usepackage{ulem}

\usepackage{color}

\begin{document}
\title{Thermally-induced vacuum instability in a single plane wave}
\author{B. \surname{King}}
\affiliation{Max-Planck-Institut f\"ur Kernphysik, Saupfercheckweg 1, D-69117 Heidelberg, Germany \\ present address: Ludwig-Maximilians-Universit\"at, Theresienstra\ss e 37, 80333 M\"unchen, Germany}
\author{H. \surname{Gies}}
\affiliation{Theoretisch-Physikalisches Institut, Friedrich-Schiller-Universit\"at Jena\\ and Helmholtz-Institut Jena, Max-Wien Platz 1, D-07743 Jena, Germany}
\author{A. \surname{Di Piazza}}
\email{dipiazza@mpi-hd.mpg.de}
\affiliation{Max-Planck-Institut f\"ur Kernphysik, Saupfercheckweg 1, D-69117 Heidelberg, Germany}

\date{\today}
\begin{abstract}
Ever since Schwinger published his influential paper [J. Schwinger, Phys. Rev. \textbf{82}, 664 (1951)], it has been unanimously accepted that the vacuum is stable in the presence of an electromagnetic plane wave. However, we advance an analysis that indicates this statement is not rigorously valid in a real situation, where thermal effects are present. We show that the thermal vacuum, in the presence of a single plane-wave field, even in the limit of zero frequency (a constant crossed field), decays into electron-positron pairs. Interestingly, the pair-production rate is found to depend nonperturbatively on both the amplitude of the constant crossed field and on the temperature.
\end{abstract}

\pacs{12.20.Ds, 11.10.Wx}
\maketitle

It has long been known that an inevitable consequence of Dirac's
theory of the electron is that in regions of sufficiently high energy
density, the quantum vacuum can break down in a spontaneous generation
of electron-positron pairs. Following the initial results of Sauter \cite{sauter31},  Heisenberg and Euler \cite{heisenberg_euler36} and Weisskopf
\cite{weisskopf36}, in a seminal work, Schwinger
\cite{schwinger51} derived a central result of strong-field quantum
electrodynamics, the rate per unit volume of pair creation $R$ in a
constant and uniform electric field of strength $E$, of leading order
behaviour $R=(E/\Ecr)^{2}(c/\lambdabar^4)(8\pi^{3})^{-1}\exp(-\pi
E_{cr}/E)$, for $E/\Ecr\ll1$, positron charge $e$, mass $m$, Compton wavelength
$\lambdabar=\hbar/mc$ and so-called ``critical'' electric field
$E_{cr} = m^{2}c^{3}/e\hbar = 1.3\times10^{16}~\trm{Vcm}^{-1}$. A
combination of factors has intensified research efforts to better
understand and devise ways of detecting this phenomenon. On the one
hand, upcoming laser facilities are planned, such as ELI (Extreme
Light Infrastructure) \cite{ELI_SDR} and HiPER (High Power laser
Energy Research) \cite{HiPER_TDR}, that intend to reach fields as
large as a percent of the critical value. On the other hand, recent
theoretical results, both analytical \cite{narozhny_creation04,
  kirk08,Bulanov10,Fedotov10,Sokolov10} and from numerical simulation
\cite{ruhl10}, strongly indicate that at orders of magnitude well
below the critical field, pair creation could be observed \cite{dipiazza12}.

Decay of the vacuum is predicted to occur in a variety of contexts. In
intense electromagnetic backgrounds such as two plane waves
propagating in different directions \cite{brezin70,dunne09,dipiazza09a} or a
plane-wave and Coulomb field combination \cite{yakovlev66}, but also
in more exotic contexts, such as being seeded by magnetic fields of
magnetars \cite{thompson08}. Another example is through thermal
radiation, with pair-creation rates having been calculated in a
constant electric field in various formalisms
\cite{gies00b,gavrilov08, kim10},
also including stimulated pair creation \cite{Monin:2008td}.
However, it has long been accepted that pair
creation can never occur in single plane waves, as encapsulated in
Schwinger's famous statement ``there are no nonlinear vacuum phenomena
for a single plane wave, of arbitrary strength and spectral
composition'' \cite{schwinger51}. The physical reason for this
statement is that all photons in a plane wave propagate in the same
direction and so cannot interact with each other. The mathematical
origin lies in the fact that the two electromagnetic invariants vanish
for a single plane wave \cite{schwinger51}.

In the current letter, we seek to demonstrate how pair creation can,
in fact, proceed, when the vacuum is polarised by a single plane wave,
if one acknowledges the inevitable presence in all real physical
scenarios, of background heat radiation. We show that this is the case
even if the frequency of the plane wave tends to zero (the so-called
``constant crossed field'' configuration). Moreover, by considering
the existence of a thermal background, we will derive an expression
for the rate of real pair creation which is non-perturbative in both
the field strength $E$ and the background temperature $T$, in both
prefactor and exponent. Apart from the conceptual advance, our results
may be of relevance for typical applications of
  the Schwinger effect such as string-breaking models
  \cite{ToporPop:2007hb} for heavy-ion collisions or for a full
  understanding of the pair production processes near neutron
  stars. In those and other cases, the thermal bath is an important
  ingredient.

We consider a scenario in which a region of the vacuum polarised by an external electromagnetic field can be considered as bathing in a thermal background, represented by a photon gas in equilibrium, at temperature $T$. The high external field intensities that interest us can be characterised with the classical non-linearity parameter $\xi = (m/\omega_{l})(E/\Ecr) \gg 1$, with $\omega_{l}$ the angular frequency of the external field photons, where we set here and subsequently $\hbar=c =k_{\text{B}}=1$. For $\xi \gg 1$, pair-creation from vacuum proceeds mainly via tunneling and in such processes, the frequency of the external field ceases to play a role, with the rate  tending to that in which the limit $\omega_{l}\to 0$ is taken \cite{ritus85}. For a plane wave, this describes the ``constant crossed field'' background, where the electric field $\bm{E}$ and the magnetic field $\bm{B}$ are constant, equal in strength ($E=B$) and perpendicular to one another. Noting that modern laser systems already allow for values of $\xi$ of the order of $10^2~$\cite{Yanovsky_2008}, we choose to study the behaviour of a photon gas in thermal equilibrium with a constant crossed field.
\begin{figure}
\begin{center}
\includegraphics[draft=false, width=0.8\linewidth]{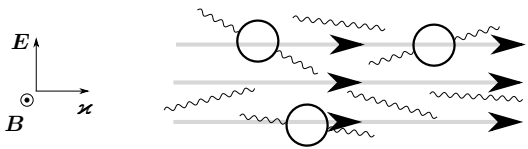}\hspace{12mm}
\end{center}
\caption{The envisaged scenario of thermal photons (wavy lines) interacting via the polarisation operator (thick circles) in equilibrium with an intense external constant crossed field $(\bm{E},\bm{B})$ (thicker gray lines, indicating the direction  $\bm{\varkappa}=\bm{E}\times\bm{B}/EB$).}\label{fig:exp_setup}
\end{figure}

The leading-order interaction between thermal photons and the external
field is contained in the polarisation operator, as displayed in
\figref{fig:exp_setup}. In general, the polarisation of the vacuum in
a background field $F^{\mu\nu}(x)$, with four-co-ordinate $x$, which
varies slowly over a Compton wavelength $\lambdabar=1/m$, can be
investigated via the effective Lagrangian of QED \cite{gies00}. It is
this object that Schwinger referred to as being identically zero for
all plane-wave backgrounds. In his analysis of a constant field and for the present case $F^{\mu\nu}(x)=F^{\mu\nu}$, there are only two relativistic invariants related to the background electromagnetic
field $F^{\mu\nu}$: $-F^{\mu\nu}F_{\mu\nu} =
2(E^{2}-B^{2})$ and
$-F^{\mu\nu}F^{\ast}_{\mu\nu}=4(\bm{E}\cdot\bm{B})$, where
$F^{\ast}_{\mu\nu}=\epsilon_{\mu\nu\alpha\beta}F^{\alpha\beta}/2$
with $\epsilon_{\mu\nu\alpha\beta}$ being a tensor antisymmetric in
all indices and $\bm{E}$ and $\bm{B}$ the electric and magnetic
field. In a plane wave, and in particular in a constant crossed field,
both the above invariants vanish identically. However, in the presence
of a thermal bath, the virtual electron-positron pairs effectively
couple the polarising background field with the thermal photons. If $k^{\mu}=(\omega,\bm{k})$ indicates
 the on-shell four-momentum of a thermal bath photon, there exists a third
relativistic invariant $(k_{\mu}F^{\mu\nu})^{2}$, which is in general non-zero in a plane-wave background. This
parameter is related to the so-called quantum non-linearity parameter
$\chi = e\sqrt{|(k_{\mu}F^{\mu\nu})^{2}|}/m^{3}$. For photons
scattering in a constant crossed field ($\bm{E},\bm{B}$) of unit
wavevector $\bm{\varkappa}=\bm{E}\times\bm{B}/EB$, we have $\chi =
\chi_{E}(\omega-\bm{k}\cdot\bm{\varkappa})/m$,
where $\chi_{E}=E/\Ecr$. Therefore, this will be the defining
microscopic variable describing the interaction between the photon gas
and the external constant crossed field. After having averaged
  over the photon vectors $k_\mu$ in the thermal bath, the additional
  relativistic invariant may also be written as $(u_\mu F^{\mu\nu})^2$,
  where $u_\mu$ is the 4-velocity vector of the heat bath. In the
  heat-bath rest frame, this invariant corresponds to $-E^2$. In the
following analysis, we limit ourselves to more accessible scenarios in
which 
$\chi$
can be regarded as much less than $1$.
Qualitatively this
occurs if the parameter $\delta=(T/m)\chi_E=(T/m)(E/\Ecr)$ is much
smaller than 1. In the following, we provide additional conditions for the validity 
of our approach. 

In regarding the thermal vacuum polarised by the external field as filled with a gas of photons, we are implicitly making two assumptions: i) that the polarisation from thermal photons can be included perturbatively and ii) the rate of pair creation from purely thermal effects is negligible. The first of these assumptions is valid if, in the formation volume of the pair production process, the number of thermal photons is less than unity. The formation length $L_{f}$, in which on average a single pair is generated, can be seen to be $L_{f}=\lambdabar/\chi_{E}$ \cite{ritus85}, whereas the density of photons in the gas is given by integrating the Bose-Einstein distribution over $d^{3}k$ to give $2\zeta(3)T^{3}/\pi^{2}$, with $\zeta(\cdot)$ the Riemann Zeta function. Therefore, the average number of thermal photons in the formation volume $L_{f}^{3}$ is $2(\zeta(3)/\pi^{2}) (\chi_E T/ m)^{3}=2(\zeta(3)/\pi^{2}) \delta^{3}$. So by fulfilling the condition $\delta^{3}\ll 1$, the first assumption will hold and in this limit, we can also obtain the total pair creation rate as an incoherent sum over the thermal ensemble. The parameters for which the second assumption is valid, can be found by first considering the case of a vacuum solely populated by a thermal photon gas. The leading order pair creation process then comes from a collision of two free photons \cite{landau4}, which can be summed over the thermal ensemble to give the background pair density $\rho_{2\gamma}(T) = dN_{2\gamma}/dVdt$ for number of pairs created $N_{2\gamma}$ with volume $V$ and time $t$. For $T/m\ll 1$ this becomes:
\be
\rho_{2\gamma}(T) \sim m^{4} \left(\frac{\alpha}{2\pi}\right)^{2} \left(\frac{T}{m}\right)^{3}\mbox{e}^{-\frac{2m}{T}},\label{R_T}
\ee
where $\alpha$ is the fine-structure constant. This background rate will later be compared to the rate of pair production arising from the interaction of the thermal bath with the constant crossed field to show where also the second assumption, and hence our analysis, is valid.

Bearing these remarks in mind, we can proceed via the optical theorem. This allows one to relate the imaginary part of the polarisation operator in an external electromagnetic field to the rate of electron-positron pair creation for a single photon propagating in this background, $R_a(k, \chi_{E}) = \textrm{Im} [l_{a}^{\mu}\Pi_{\mu\nu}(k, \chi_{E})l_{a}^{*\nu}]/\omega$, for normalised photon four-polarisation $l_{a}^{\mu}$ ($\sum_a l_{a}^{\mu}l_{a}^{*\nu}\to -g^{\mu\nu}$), $a=\{1,2\}$ and polarisation operator $\Pi_{\mu\nu}(k, \chi_{E})$. For brevity, we insert the derived polarisation operator in a constant crossed field background (see e.g. \cite{baier75a}) into this expression, giving a single-photon rate:
\begin{equation}
\label{eqn:rate1}
R_a(k, \chi_{E})=-\frac{2\alpha m^{2}}{3\pi\omega}\!\int^{\infty}_{4}\!\!dv \frac{2v+1+3(-1)^a }{v\sqrt{v(v-4)}}\frac{\trm{Ai}^{\prime}(z)}{z},
\end{equation}
where $z=(v/\chi)^{2/3}$, $\trm{Ai}(\cdot)$ is the Airy function of the first kind \cite{olver97}.

In the situation of interest where the value of the parameter $\delta$ is very small, we can regard the condition $\chi\ll1$ to be satisfied by all save a negligible number of photons. Therefore, we can asymptotically expand \eqnref{eqn:rate1}, which will be further justified in the discussion, to give:
\be
R_a(k, \chi_{E}) \sim \alpha m \frac{3+(-1)^a }{16}\sqrt\frac{3}{2}\,\frac{m}{\omega}\chi \mbox{e}^{-\frac{8}{3\chi}}.\label{eqn:rate2}
\ee
The density of pairs created by the external field-thermal bath interaction $\rho_{\trm{th}}(T, \chi_{E}) = dN_{\trm{th}}/dVdt$, can then be obtained by summing \eqnref{eqn:rate2} over the Bose-Einstein distribution, well approximated by an integration:
\be
\rho_{\trm{th}}(T,\chi_{E}) = \sum_{a}\int \frac{d^{3}k}{(2\pi)^3}\, \frac{1}{\mbox{e}^{\omega/T}-1}\, R_a(k, \chi_{E}).\label{eqn:rate3}
\ee
We mention at this point that since our analysis assumes a thermal equilibrium, as soon as the first pair is created, the conditions under which our result is valid, are altered. Therefore, \eqnref{eqn:rate3} should be understood in the sense of a probability per unit volume per unit time for pair creation to ensue, rather than a rate per unit volume.

\begin{figure}
\begin{center}
\includegraphics[width=\linewidth]{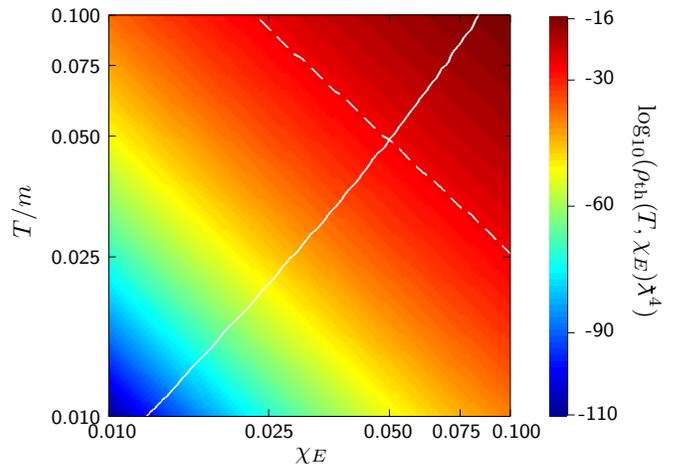}
\end{center}
\caption{Logarithm of the expected number of pairs generated in a constant crossed field in the tiny space-time volume $\lambdabar^{4} = 7.4\times 10^{-53}~\trm{cm}^{3}\,\trm{s}$. The area above the solid line signifies where the background process of pair creation due to purely thermal effects (see Eq. (\ref{R_T})) is at least ten percent of that due to the external field interacting with the thermal photon gas (see Eq. (\ref{eqn:rate5})). Above the 
dashed line, the number of pairs created in a typical strong optical laser four-volume $\Omega_l=6\times 10^{-26}~\trm{cm}^{3}\trm{s}$, is larger than unity.}\label{fig:rate}
\end{figure}

Inserting the asymptotic rate \eqnref{eqn:rate2} into \eqnref{eqn:rate3}, making the substitutions $\eta=\omega/T$ and $y=1-\cos\theta$, with $\cos\theta=\bm{k}\cdot\bm{\varkappa}/\omega$ and noticing that at small $\chi$, only large values of $\eta$ contribute to the integral, and we arrive at:
\be
\rho_{\trm{th}}(T,\chi_{E})\sim\frac{3\sqrt{3}T^{3}\alpha m \chi_{E}}{32\sqrt{2}\pi^{2}}\!\!\int^{\infty}_{0}\!\!\!\!d\eta\!\! \int^{2}_{0}\!\!dy \,\eta^{2}y \frac{\mbox{e}^{-\eta}}{\mbox{e}^{8/3\eta y\delta}}.\label{eqn:rate4}
\ee
The remaining integrations can be also performed, giving the final result:
\be
\rho_{\trm{th}}(T,\chi_E) \sim \frac{3^{3/4}\alpha }{4\pi^{3/2}}\,m^{4}\,\left(\frac{T}{m}\right)^{2}\delta^{1/4} \mbox{e}^{-\frac{4}{\sqrt{3\delta}}}, \,\, \delta=\frac{T}{m}\chi_{E}. \label{eqn:rate5} 
\ee
This result is valid if $\sqrt{\delta}\ll 1$ and it is interesting for several reasons. The rate shows a non-perturbative dependency on both the temperature and the external field, in both its pre-exponent and exponent and in a way quite different from the classic Schwinger result quoted in the introduction. In \figref{fig:rate} we plot the logarithm of the rate $\rho_{\trm{th}}(T,\chi_{E})$, calculated from \eqnref{eqn:rate5}, times the four-dimensional volume $\lambdabar^{4}=1/m^4 = 7.4\times10^{-53}~\trm{cm}^{3}\,\trm{s}$. In the region above the 
dashed line, the number of pairs created in a typical optical strong laser beam four-volume $\Omega_{l}=\tau_{l}V_{l}$, where $\tau_{l}=10~\trm{fs}$, $V_{l}=\pi\times(0.8\,\mu\trm{m})^{2}\times c\tau_{l}=6\times 10^{-12}~\trm{cm}^{3}$ (where we temporarily recover the speed of light, $c$), is larger than one. Upon inspection of \figref{fig:rate}, we notice that for high enough temperatures, even with $E=0.025$, an exponential number of pairs can be produced in a typical optical laser pulse. This should be compared to the case of zero temperature, for which the expected number of pairs is identically zero. The solid line in \figref{fig:rate} shows the significance of the background process of pure thermal pair creation (see Eq. (\ref{R_T})), which dominates for large $T/m$ and small $\chi_{E}$. By comparing the analytical expressions in Eqs. (\ref{R_T}) and (\ref{eqn:rate5}), we obtain the condition that the pure thermal pair production process is negligible if $\alpha(T/m)\exp(-2m/T)\ll\delta^{1/4} \exp(-4/\sqrt{3\delta})$. The simpler but over-conservative condition for the validity of our method, $T/m\ll\sqrt{\delta}\ll1$, is obtained by comparing the exponents in Eqs. (\ref{R_T}) and (\ref{eqn:rate5}).

The physical line of argument given above also has a formal
 counterpart that yields additional insight into our result
  \eqref{eqn:rate5}. Averaging the one-loop polarization tensor over a
  thermal photon bath corresponds exactly to a two-loop calculation
  involving a zero-temperature electron-positron loop and a
  thermalised photon radiative correction. In the finite-temperature 
Matsubara formalism, if one makes the assumption that only the photon propagator 
need be thermalised (justified below), the rate is given by:
\be
\rho_{\trm{M}}(T,\chi_{E}) = -T\,\trm{Im}\int \frac{d^{3}k}{(2\pi)^{3}} \sum_{n=-\infty}^{\infty}\frac{1}{k_{n}^{2}}~g_{\mu\nu}\Pi^{\mu\nu}(k_{n}, \chi_{E}), \label{eqn:Matsu1}
\ee
where $k_{n}=(2 \pi i nT, \bm{k})$.
By using an integral relation to transform the sum into a complex integral (see Eq. (4.11) in \cite{quiros07}), 
one can show that \eqnref{eqn:Matsu1} and the thermal average \eqnref{eqn:rate3}, coincide.
The two-loop
  effective action has been calculated in the Matsubara formalism for
  general constant-field backgrounds in \cite{gies00b, gies00} in the form of a
  triple propertime-integral representation also involving a sum over
  the windings of the photon around the Euclidean finite-$T$
  cylinder. Taking the crossed-field limit, the result
  \eqref{eqn:rate5} arises from a saddle point of the propertime
  integrals, which occurs under the condition that
  $\gamma_T=T^{2}/m^{2}\delta\ll 1$. 
  The requirement of small $\gamma_T$ is not only in line with the
  assumptions made above, but also shows that our pair production
  density is dominated by non-perturbative Schwinger-type pair
  production in contrast to a multi-photon regime for large
  $\gamma_T$.

Another lesson to be learned from the effective action
representation is that pair production in crossed fields is
triggered by thermal photons with high winding number $n$ in
Euclidean space. Contributions to pair production arise from
winding numbers $n>2\sqrt{3}\,\gamma_T$. In the $\gamma_T\ll 1$
limit this is satisfied for all $n\geq 1$. In the general case,
it shows that pair production requires a minimum amount of
delocalization of the virtual electron-positron pairs in order to
acquire sufficient energy to become real. The winding number $n$
is a measure for this delocalization.

Finally, we can also extract information about the fundamental
validity limit of the above reasoning. For increasing
temperature, also the thermal fluctuations of the
electron-positrons need to be taken into account. For $T\ll m$,
they are typically exponentially suppressed $\sim \exp(-m/T)$
\cite{gies00, Elmfors:1994fw, gies99}, such that we expect our approximation of
thermalising only the photon to hold for $T\lesssim m$.

Our results also confirm the tendency observed for Schwinger
  pair production that the thermal contribution exceeds the vacuum
  contribution in the limit of weak fields \cite{gies00}. In the
  present crossed-field case, this is particularly evident, as the
  vacuum production rate in crossed fields is exactly zero.

\begin{figure}
\begin{center}
\includegraphics[width=0.68\linewidth]{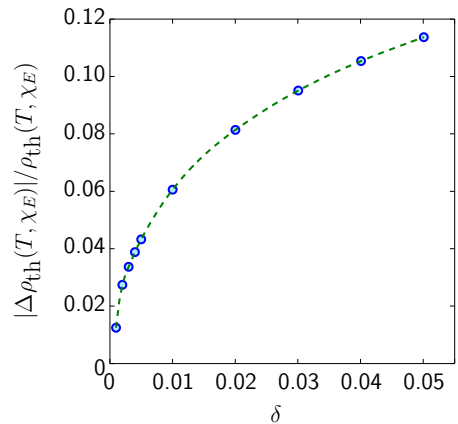}
\end{center}
\caption{Relative difference $|\Delta\rho_{\trm{th}}(T,\chi_{E})|/\rho_{\trm{th}}(T,\chi_{E})$ between the exact and asymptotic formulae for the thermal density of created pairs (using \eqnref{eqn:rate1} in Eq. (\ref{eqn:rate3}) and using Eq. (\ref{eqn:rate5}), respectively), plotted as a function of $\delta$.} \label{fig:asymptotic_plot}
\end{figure}

In order to justify the validity of using the asymptotic limit for single-photon pair creation, we can numerically calculate the ensemble pair density by inserting the full single-photon rate \eqnref{eqn:rate1} into the thermal integral \eqnref{eqn:rate3} and compare it with the analytical expression in Eq. (\ref{eqn:rate5}). 
Upon doing so, we see that up to values of $\delta=0.03$ (the highest value of $\delta$ plotted in \figref{fig:rate} is $0.01$), there is a maximum deviation of around $10\%$, as plotted in \figref{fig:asymptotic_plot}. Therefore, the asymptotic expression for the created pair density in \eqnref{eqn:rate5} gives a good approximation to the exact value calculated by this method. We note that by passing to the integration variable $\eta=\omega/T$ in \eqnref{eqn:rate3}, it can be shown that the rate $\rho_{\trm{th}}$ has the form $\rho_{\trm{th}}(T,\chi_{E})=(T/m)^{2}\widetilde{\rho}_{\trm{th}}(\delta)$ such that the explicit dependence on $T/m$ cancels in $|\Delta\rho_{\trm{th}}(T,\chi_{E})|/\rho_{\trm{th}}(T,\chi_{E})$.

We conclude by recapitulating the main results of the letter. We have shown how, in a region of the thermal vacuum in equilibrium with an external constant crossed field, pair creation can indeed take place. By extension, this conclusion also applies to regions of the vacuum polarised by single plane waves, when in the tunneling regime. The probability for this new process is non-perturbative in both the external field strength and the temperature of the thermal vacuum and is distinct from other analytical results for pair creation by external fields. For realistic field intensities orders of magnitude less than critical, a thermal background can seed the creation of an exponential number of particles. Thus the old adage that the vacuum cannot be polarised by a single plane wave, although useful in understanding the theoretical basis of vacuum polarisation, is an idealisation which yields when heat and hence any physically realistic scenario, is taken into account.

H.~G. acknowledges support by the DFG under grants SFB-TR18 and
GI~328/4-1 (Heisenberg program).

\providecommand{\noopsort}[1]{}


\begin{thebibliography}{28}
\expandafter\ifx\csname natexlab\endcsname\relax\def\natexlab#1{#1}\fi
\expandafter\ifx\csname bibnamefont\endcsname\relax
  \def\bibnamefont#1{#1}\fi
\expandafter\ifx\csname bibfnamefont\endcsname\relax
  \def\bibfnamefont#1{#1}\fi
\expandafter\ifx\csname citenamefont\endcsname\relax
  \def\citenamefont#1{#1}\fi
\expandafter\ifx\csname url\endcsname\relax
  \def\url#1{\texttt{#1}}\fi
\expandafter\ifx\csname urlprefix\endcsname\relax\def\urlprefix{URL }\fi
\providecommand{\bibinfo}[2]{#2}
\providecommand{\eprint}[2][]{\url{#2}}

\bibitem[{\citenamefont{Sauter}(1931)}]{sauter31}
\bibinfo{author}{\bibfnamefont{F.}~\bibnamefont{Sauter}}, \bibinfo{journal}{Z.
  Phys.} \textbf{\bibinfo{volume}{69}}, \bibinfo{pages}{742}
  (\bibinfo{year}{1931}).

\bibitem[{\citenamefont{Heisenberg and Euler}(1936)}]{heisenberg_euler36}
\bibinfo{author}{\bibfnamefont{W.}~\bibnamefont{Heisenberg}} \bibnamefont{and}
  \bibinfo{author}{\bibfnamefont{H.}~\bibnamefont{Euler}}, \bibinfo{journal}{Z.
  Phys.} \textbf{\bibinfo{volume}{98}}, \bibinfo{pages}{714}
  (\bibinfo{year}{1936}).

\bibitem[{\citenamefont{Weisskopf}(1936)}]{weisskopf36}
\bibinfo{author}{\bibfnamefont{V.}~\bibnamefont{Weisskopf}},
  \bibinfo{journal}{Mat.-fys. Medd.} \textbf{\bibinfo{volume}{14}},
  \bibinfo{pages}{6} (\bibinfo{year}{1936}).

\bibitem[{\citenamefont{Schwinger}(1951)}]{schwinger51}
\bibinfo{author}{\bibfnamefont{J.}~\bibnamefont{Schwinger}},
  \bibinfo{journal}{Phys. Rev.} \textbf{\bibinfo{volume}{82}},
  \bibinfo{pages}{664} (\bibinfo{year}{1951}).

\bibitem[{ELI()}]{ELI_SDR}
\bibinfo{howpublished}{http://www.extreme-light-infrastructure.eu}.

\bibitem[{HiP()}]{HiPER_TDR}
\bibinfo{howpublished}{http://www.hiperlaser.org}.

\bibitem[{\citenamefont{Narozhny et~al.}(2004)}]{narozhny_creation04}
\bibinfo{author}{\bibfnamefont{N.~B.} \bibnamefont{Narozhny}}
  \bibnamefont{et~al.}, \bibinfo{journal}{Phys. Lett. A}
  \textbf{\bibinfo{volume}{330}}, \bibinfo{pages}{1} (\bibinfo{year}{2004}).

\bibitem[{\citenamefont{Bell and Kirk}(2008)}]{kirk08}
\bibinfo{author}{\bibfnamefont{A.~R.} \bibnamefont{Bell}} \bibnamefont{and}
  \bibinfo{author}{\bibfnamefont{J.~G.} \bibnamefont{Kirk}},
  \bibinfo{journal}{Phys. Rev. Lett.} \textbf{\bibinfo{volume}{101}},
  \bibinfo{pages}{200403} (\bibinfo{year}{2008}).

\bibitem[{\citenamefont{Bulanov et~al.}(2010)}]{Bulanov10}
\bibinfo{author}{\bibfnamefont{S.~S.} \bibnamefont{Bulanov}}
  \bibnamefont{et~al.}, \bibinfo{journal}{Phys. Rev. Lett.}
  \textbf{\bibinfo{volume}{105}}, \bibinfo{pages}{220407}
  (\bibinfo{year}{2010}).

\bibitem[{\citenamefont{Fedotov et~al.}(2010)}]{Fedotov10}
\bibinfo{author}{\bibfnamefont{A.~M.} \bibnamefont{Fedotov}}
  \bibnamefont{et~al.}, \bibinfo{journal}{Phys. Rev. Lett.}
  \textbf{\bibinfo{volume}{105}}, \bibinfo{pages}{080402}
  (\bibinfo{year}{2010}).

\bibitem[{\citenamefont{Sokolov et~al.}(2010)}]{Sokolov10}
\bibinfo{author}{\bibfnamefont{I.~V.} \bibnamefont{Sokolov}}
  \bibnamefont{et~al.}, \bibinfo{journal}{Phys. Rev. Lett.}
  \textbf{\bibinfo{volume}{105}}, \bibinfo{pages}{195005}
  (\bibinfo{year}{2010}).

\bibitem[{\citenamefont{Nerush et~al.}(2011)}]{ruhl10}
\bibinfo{author}{\bibfnamefont{E.~N.} \bibnamefont{Nerush}}
  \bibnamefont{et~al.}, \bibinfo{journal}{Phys. Rev. Lett.}
  \textbf{\bibinfo{volume}{106}}, \bibinfo{pages}{035001}
  (\bibinfo{year}{2011}).

\bibitem[{\citenamefont{Di Piazza et~al.}(2012)}]{dipiazza12}
\bibinfo{author}{\bibfnamefont{A.~Di} \bibnamefont{Piazza}}
  \bibnamefont{et~al.}, \bibinfo{howpublished}{http://arXiv:hep-ph/1111.3886v1}
  (\bibinfo{year}{2012}).


\bibitem{brezin70}
E.~Br\'ezin and C.~Itzykson, Phys.\ Rev.\ D {\bf 2}, 1191 (1970); 
V.~S.~Popov, JETP\ Lett.\ {\bf 13}, 185 (1971). 


\bibitem[{\citenamefont{Dunne et~al.}(2009)\citenamefont{Dunne, Gies, and
  Sch{\"u}tzhold}}]{dunne09}
\bibinfo{author}{\bibfnamefont{G.~V.} \bibnamefont{Dunne}},
  \bibinfo{author}{\bibfnamefont{H.}~\bibnamefont{Gies}}, \bibnamefont{and}
  \bibinfo{author}{\bibfnamefont{R.}~\bibnamefont{Sch{\"u}tzhold}},
  \bibinfo{journal}{Phys. Rev. D} \textbf{\bibinfo{volume}{80}},
  \bibinfo{pages}{111301} (\bibinfo{year}{2009}).



\bibitem[{\citenamefont{Di Piazza et~al.}(2009)\citenamefont{Di Piazza}}]{dipiazza09a}
\bibinfo{author}{\bibfnamefont{A.~Di} \bibnamefont{Piazza}},
\bibnamefont{et~al.}, \bibinfo{journal}{Phys. Rev. Lett.} \textbf{\bibinfo{volume}{103}},
  \bibinfo{pages}{170403} (\bibinfo{year}{2009}).


\bibitem[{\citenamefont{Yakovlev}(1966)}]{yakovlev66}
\bibinfo{author}{\bibfnamefont{V.~P.} \bibnamefont{Yakovlev}},
  \bibinfo{journal}{Sov. Phys. JETP} \textbf{\bibinfo{volume}{22}},
  \bibinfo{pages}{223} (\bibinfo{year}{1966}).

\bibitem[{\citenamefont{Thompson}(2008)}]{thompson08}
\bibinfo{author}{\bibfnamefont{C.}~\bibnamefont{Thompson}},
  \bibinfo{journal}{Astrophys. J.} \textbf{\bibinfo{volume}{2}},
  \bibinfo{pages}{1258} (\bibinfo{year}{2008}).

\bibitem[{\citenamefont{Gies}(2000)}]{gies00b}
\bibinfo{author}{\bibfnamefont{H.}~\bibnamefont{Gies}}, \bibinfo{journal}{Phys.
  Rev. D} \textbf{\bibinfo{volume}{61}}, \bibinfo{pages}{085021}
  (\bibinfo{year}{2000}).

\bibitem[{\citenamefont{Gavrilov and Gitman}(2008)}]{gavrilov08}
\bibinfo{author}{\bibfnamefont{S.~P.} \bibnamefont{Gavrilov}} \bibnamefont{and}
  \bibinfo{author}{\bibfnamefont{D.~M.} \bibnamefont{Gitman}},
  \bibinfo{journal}{Phys. Rev. D} \textbf{\bibinfo{volume}{78}},
  \bibinfo{pages}{045017} (\bibinfo{year}{2008}).

\bibitem[{\citenamefont{Kim et~al.}(2010)\citenamefont{Kim, Lee, and
  Yoon}}]{kim10}
\bibinfo{author}{\bibfnamefont{S.~P.} \bibnamefont{Kim}},
  \bibinfo{author}{\bibfnamefont{H.~K.} \bibnamefont{Lee}}, \bibnamefont{and}
  \bibinfo{author}{\bibfnamefont{Y.}~\bibnamefont{Yoon}},
  \bibinfo{journal}{Phys. Rev. D} \textbf{\bibinfo{volume}{82}},
  \bibinfo{pages}{025016} (\bibinfo{year}{2010}).

\bibitem{Monin:2008td} A.~K.~Monin and A.~V.~Zayakin, JETP Lett.\  {\bf 87}, 709 (2008).

\bibitem{ToporPop:2007hb}  V.~Topor Pop et~al., Phys.\ Rev.\ C {\bf 75}, 014904 (2007).

\bibitem[{\citenamefont{Ritus}(1985)}]{ritus85}
\bibinfo{author}{\bibfnamefont{V.~I.} \bibnamefont{Ritus}},
  \bibinfo{journal}{J. Russ. Laser Res.} \textbf{\bibinfo{volume}{6}},
  \bibinfo{pages}{497} (\bibinfo{year}{1985}).


\bibitem[{\citenamefont{Yanovsky et~al.}(2008)}]{Yanovsky_2008}
\bibinfo{author}{\bibfnamefont{V.}~\bibnamefont{Yanovsky}}
  \bibnamefont{et~al.}, \bibinfo{journal}{Opt. Express}
  \textbf{\bibinfo{volume}{16}}, \bibinfo{pages}{2109} (\bibinfo{year}{2008}).

\bibitem[{\citenamefont{Dittrich and Gies}(2000)}]{gies00}
\bibinfo{author}{\bibfnamefont{W.}~\bibnamefont{Dittrich}} \bibnamefont{and}
  \bibinfo{author}{\bibfnamefont{H.}~\bibnamefont{Gies}},
  \textit{\bibinfo{title}{Probing the Quantum Vacuum}}
  (\bibinfo{publisher}{Springer-Verlag}, \bibinfo{address}{Berlin},
  \bibinfo{year}{2000}).

\bibitem[{\citenamefont{Berestetskii et~al.}(1982)\citenamefont{Berestetskii,
  Lifshitz, and Pitaevskii}}]{landau4}
\bibinfo{author}{\bibfnamefont{V.~B.} \bibnamefont{Berestetskii}},
  \bibinfo{author}{\bibfnamefont{E.~M.} \bibnamefont{Lifshitz}},
  \bibnamefont{and} \bibinfo{author}{\bibfnamefont{L.~P.}
  \bibnamefont{Pitaevskii}}, \textit{\bibinfo{title}{Quantum Electrodynamics}}
  (\bibinfo{publisher}{Butterworth-Heinemann}, \bibinfo{address}{Oxford},
  \bibinfo{year}{1982}).

\bibitem[{\citenamefont{Ba\u{\i}er et~al.}(1976)\citenamefont{Ba\u{\i}er,
  Mil'shte\u{\i}n, and Strakhovenko}}]{baier75a}
\bibinfo{author}{\bibfnamefont{V.~N.} \bibnamefont{Ba\u{\i}er}},
  \bibinfo{author}{\bibfnamefont{A.~I.} \bibnamefont{Mil'shte\u{\i}n}},
  \bibnamefont{and} \bibinfo{author}{\bibfnamefont{V.~M.}
  \bibnamefont{Strakhovenko}}, \bibinfo{journal}{Sov. Phys. JETP}
  \textbf{\bibinfo{volume}{42}}, \bibinfo{pages}{961} (\bibinfo{year}{1976}).

\bibitem[{\citenamefont{Olver}(1997)}]{olver97}
\bibinfo{author}{\bibfnamefont{F.~W.~J.} \bibnamefont{Olver}},
  \textit{\bibinfo{title}{Asymptotics and Special Functions}}
  (\bibinfo{address}{AKP Classics, Natick}, \bibinfo{year}{1997}).

\bibitem[{\citenamefont{Quir{\'o}s}(2007)}]{quiros07}
\bibinfo{author}{\bibfnamefont{M.}~\bibnamefont{Quir{\'o}s}}, \bibinfo{journal}{Acta.
 Phys. Pol. B} \textbf{\bibinfo{volume}{38}}, \bibinfo{pages}{3661}
  (\bibinfo{year}{2007}).


\bibitem{Elmfors:1994fw} P.~Elmfors and B.~-S.~Skagerstam, Phys.\ Lett.\ B {\bf 348}, 141 (1995) [Erratum-ibid.\ B {\bf 376}, 330 (1996)]

\bibitem[{\citenamefont{Gies}(1999)}]{gies99}
\bibinfo{author}{\bibfnamefont{H.}~\bibnamefont{Gies}}, \bibinfo{journal}{Phys.
  Rev. D} \textbf{\bibinfo{volume}{60}}, \bibinfo{pages}{105002}
  (\bibinfo{year}{1999}).

\end{thebibliography}
\end{document}